\documentclass[a4paper]{jpconf}

\usepackage{enumitem}
\usepackage{graphicx}
\usepackage{color}

\usepackage[pdftex,colorlinks=true,bookmarks=false,citecolor=blue,urlcolor=blue]{hyperref} 

\begin{document}

\title{Status of advanced ground-based laser interferometers for gravitational-wave detection}

\author{K~L~Dooley$^1$ \footnote{Current address: The University of Mississippi, University, MS 38677, USA}, T~Akutsu$^2$, S~Dwyer$^3$, P~Puppo$^4$}
\address{$^1$ Max-Planck-Institut f\"ur Gravitationsphysik
(Albert-Einstein-Institut) und Leibniz Universit\"at Hannover, Callinstr. 38,
D-30167 Hannover, Germany.}
\address{$^2$ National Astronomical Observatory of Japan, 2-21-1 Osawa, Mitaka, Tokyo 181-8588} 
\address{$^3$ LIGO Hanford Observatory, PO Box 159, Richland, WA 99352, USA}
\address{$^4$ INFN, Sezione di Roma, I-00185 Roma, Italy}

\ead{kate.dooley@aei.mpg.de}

\date{\today}

\begin{abstract}
Ground-based laser interferometers for gravitational-wave (GW)
detection were first constructed starting 20 years ago and as of 2010
collection of several years' worth of science data at initial design
sensitivities was completed. Upgrades to the initial detectors
together with construction of brand new detectors are ongoing and
feature advanced technologies to improve the sensitivity to GWs. This
conference proceeding provides an overview of the common design
features of ground-based laser interferometric GW detectors and
establishes the context for the status updates of each of the four
gravitational-wave detectors around the world: Advanced LIGO, Advanced
Virgo, GEO\,600 and KAGRA.
\end{abstract}

\section{Introduction}
This is an exciting time for the maturing field of gravitational wave
(GW) physics. The network of ground-based laser interferometer GW
detectors (GWD) is making rapid progress towards its goal of producing
advanced instruments sensitive enough to make monthly GW detections in
the 10\,Hz to 10\,kHz frequency range by the end of the
decade \cite{Abadie2010Predictions}. Simultaneously, there has
recently been a claim by the BICEP2 experiment that the measurement of
the B-modes in the cosmic microwave background polarization can be the
signature of the primordial GWs produced by
inflation \cite{Ade2014}. If confirmed, this would add to the original
evidence for GWs from 1975 by Hulse and Taylor who observed that the
rate of change of the orbital period of a binary star system precisely
agrees with the predictions of GR \cite{Hulse1975Discovery,
Weisberg2005Relativistic}. Moreover, the eLISA space-based
gravitational wave detector \cite{Danzmann1996} together with the
promising pulsar timing technique \cite{Hobbs2010International} will
be very important in the coming decades for the investigation of very
massive objects and other GW sources in the milli- to micro-Hz
frequency ranges, respectively.

GWs are dynamic strains in space-time that travel at the speed of
light and are generated by non-axisymmetric acceleration of mass. They
perturb the flat Minkowski metric describing space-time. The effect is
the production of a dimensionless strain between two inertial masses
located at a proper distance $L$ from one another so that their distance
changes as:
\begin{equation}
\Delta L(t)=\frac{1}{2} ~ L ~ h(t)
\end{equation}
In the late 20th century following the era of the operation of bar
detectors for GW detection \cite{Weber1960Detection}, large laser
interferometers were identified as the promising route forward because
of the very high strain sensitivities that could be achieved over a
wide frequency band. R.~Weiss produced the first detailed design study
in 1972 of a large scale laser interferometer for GW detection,
complete with calculations of fundamental noise sources
\cite{WeissElectromagnetically}. Then, following the work of Forward
\cite{Forward1978Wideband} who built the first laser interferometric
GW detector prototype, many groups around the world proceeded to study
the benefits of laser interferometry, build new prototypes, perfect
the design, and push technology development \cite{Shoemaker1988Noise,
  Robertson1995Glasgow, Livas1985MIT, Abramovici1996Improved}.

\begin{figure}[Htb]
\centering
\includegraphics[width=0.8\textwidth]{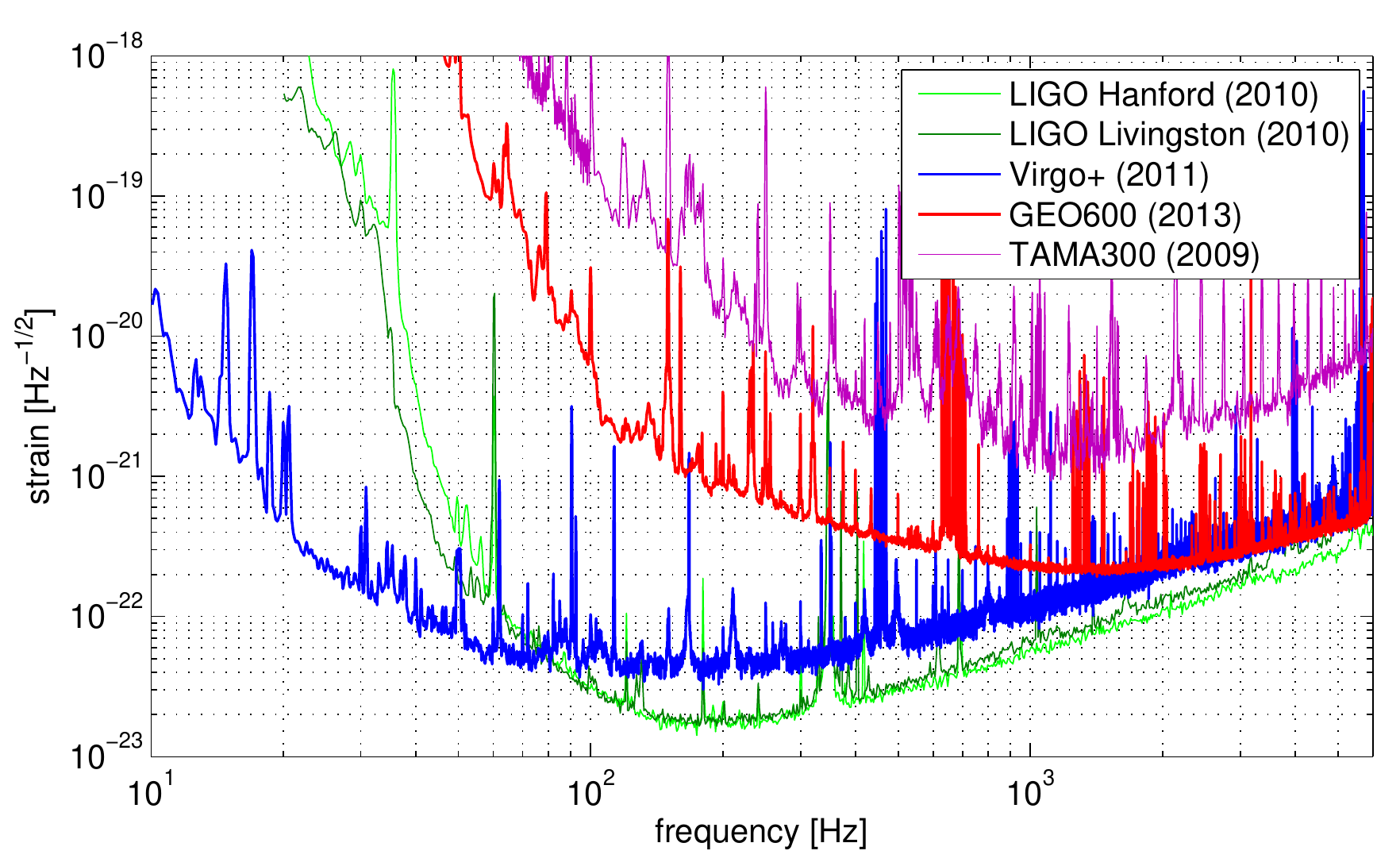}
\caption{Strain sensitivities achieved by the first generation detectors.}
\label{fig:sensitivity}
\end{figure}

The LIGO detectors in the U.S. \cite{Abbott2009LIGO}, the Virgo
detector in Italy \cite{Acernese2008Virgo}, the GEO\,600 detector in
Germany \cite{Grote2010GEO} and the TAMA\,300~\cite{Takahashi04}
detector in Japan formed the first generation laser-interferometric
GWD network. Construction of these projects began in the mid-1990s and
then progressed in sequential commissioning and data-taking phases in
the 2000s. These first-generation detectors achieved their original
instrument sensitivity goals and are now undergoing major hardware
upgrades to expand thousand-fold the observable volume of the
cosmos. The strain sensitivities achieved by these initial detectors
is shown in Figure~\ref{fig:sensitivity}. This paper describes the
common design basics of the detectors and provides an overall summary
of the current status of the worldwide network of GWDs, including a
timeline for operations. The next four proceedings focus on the
particular features and individual status of each of the advanced
versions of these detectors \cite{Dwyer2014LIGO, Puppo2014Virgo,
Dooley2014GEO600, Akutsu2014Kagra}.
 
\section{Background}
A direct detection of gravitational waves themselves has not yet been
made, but this is also not a surprise. The rate estimates for
coalescing binary neutron stars, for instance, predict a detection
probability of one event in 100 years for the initial detector
sensitivities \cite{Abadie2010Predictions}. About two years' worth of
double-coincidence data were collected. Nonetheless, the LIGO and Virgo
collaborations have already produced astrophysical results from the
data collected thus far. They have placed an upper limit that beats
previous best estimates of the fraction of spin-down power emitted in
GWs from the Crab Pulsar \cite{Abbott2008Beating}. They have also placed
an upper limit at 100\,Hz on the normalized energy density of the
stochastic GW background of cosmological and astrophysical origin, a
result otherwise inaccessible to standard observational techniques
\cite{S5NatureStochastic}.

The last decade brought great advances in demonstrating the
experimental feasibility of achieving the strain sensitivities
required to witness astrophysical events and informed the design of
today's generation of GWDs. Both LIGO and Virgo are re-using the
infrastructure from the initial generation detectors and are replacing
the hardware within the vacuum system. The upgraded detectors are
called Advanced LIGO \cite{Harry2010Advanced} and Advanced
Virgo \cite{AdVTDR}, respectively. TAMA\,300 as well as CLIO, a
prototype cryogenic laser interferometer \cite{Uchiyama2012Reduction},
informed the design of a brand new detector called KAGRA, the first
underground, cryogenic laser interferometer \cite{Somiya2012Detector,
Arai2009Status}. GEO\,600 is keeping its infrastructure and most of
its initial generation hardware, but is carrying out upgrades to
demonstrate advanced techniques in a program called
GEO-HF \cite{Luck2010Upgrade}. In addition, a proposal to expand
LIGO's baseline by building an interferometer in India is moving
forward.

\section{Detector design overview}
\label{overview}

\begin{figure}[tb]
\centering
\scalebox{0.8}{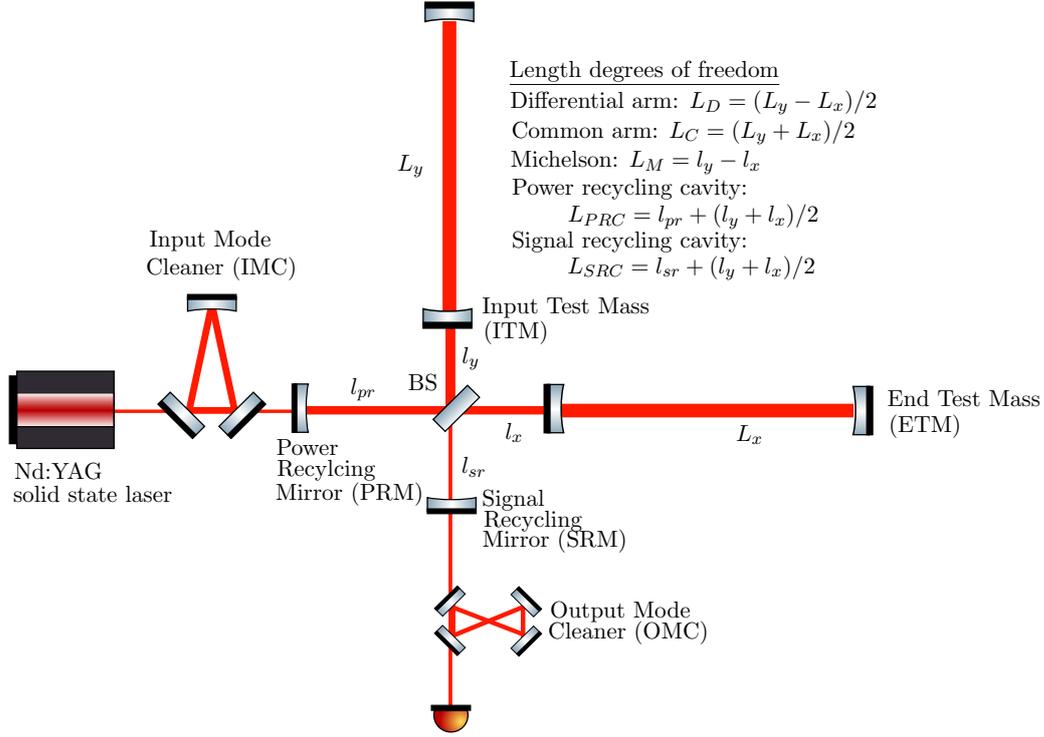}  
\caption{Layout of advanced ground-based interferometric gravitational
  wave detectors (not to scale). To keep the interferometer in its
  linear operating range, the length sensing and control system must
  control five interferometer length degrees of freedom and keep both
  mode cleaner cavities on resonance. The differential arm degree of
  freedom is sensitive to gravitational waves, and is sensed by the
  photo-detector in transmission of the output mode cleaner. GEO\,600
  is the exception in that it has no arm cavities.}
\label{fig:IFOschematic}
\end{figure}

The typical advanced detector configuration is that of a dual-recycled
Fabry-Perot Michelson (DRFPM) laser interferometer as depicted in
Figure~\ref{fig:IFOschematic}. A power amplified, and intensity and
frequency stabilized Nd:YAG solid state laser system injects
linear-polarized 1064~nm light into a triangular input mode-cleaner
(IMC) cavity. The IMC suppresses laser frequency noise and provides
spatial filtering of the laser beam to reduce beam jitter that could
otherwise couple to the GW readout. A beam splitter (BS) sends the
beam to the two Fabry-Perot arms, which are made of an input test mass
mirror (ITM) and an end test mass mirror (ETM). Both arms are of
km-scale lengths and are set to maintain nearly perfect destructive
interference of the recombined light at the anti-symmetric (AS) port
which carries the GW information. Here, the beam is directed to an
output mode-cleaning (OMC) cavity and then onto a photo-detector. The
OMC transmits only the signal-carrying light to improve the
signal-to-noise ratio.

A power recycling mirror at the symmetric port directs the
constructively-interfered light back into the interferometer. The
transmissivity of the power recycling mirror is set to match the
losses of the main optics to create a nearly critically coupled
cavity. A signal recycling mirror at the anti-symmetric port creates
an additional cavity which can be used to adjust the storage time of
the gravitational wave signal in the interferometer and thus the
frequency response of the detector. The signal recycling mirror has a
transmissivity selected to compromise between high and low frequency
sensitivity based on thermal noise and laser power. Until a few years
ago all detectors used heterodyne readout. Since approximately 2008 as
part of intermediary upgrades to the initial detectors
\cite{Adhikari2006Enhanced}, homodyne (\emph{DC}) readout was
implemented together with the addition of an OMC \cite{Fricke2012DC,
  Hild2009DCreadout}.

Nearly the entire interferometer is enclosed in an ultra high vacuum
(UHV) system to render phase noise of residual gas unimportant and to
keep the optics free of dust. The primary interferometer optics are
suspended as pendulums to decouple them from ground motion so that
they act like free masses in the horizontal plane at the frequencies
in the GW detection band. To minimize the impact of thermal noise, the
mirror suspensions are designed to minimize dissipation and, in the
case of KAGRA, operated at cryogenic temperatures. Each mirror is
equipped with actuators for coarse and fine control of the mirror
position and orientation.

A feedback control system is implemented to hold the system
sufficiently near the intended operating point so that the response to
residual deviations remains linear. Calibration of the detector must
take into account the action of the control system and the frequency
response of the detector \cite{Hewitson2004Principles}.
The various length (illustrated in Figure \ref{fig:IFOschematic}) and
angular degrees of freedom are sensed through the use of
radio-frequency sidebands on the carrier light that are created
through phase modulation by electro-optic modulators. The differential
arm length signal is sensitive to gravitational waves, and is sensed
using homodyne readout in transmission of the OMC, where the GW signal
is encoded as power variations of the light. The standard technique
for locking optical cavities is the Pound--Drever--Hall method of
laser frequency stabilization \cite{Drever1983}. Although the
interferometer is an analog instrument, it is interfaced through a
digital control system which allows complex filters to be implemented
and tuned from the control room.

The sources of noise that contaminate the detector's output can be
grouped into two categories: displacement noise and sensing
noise. Displacement noises are those that create real motion of the
mirrors, while sensing noises are those that arise in the process of
measuring the mirrors' position. The primary displacement noise that
plagues terrestrial laser interferometers at very low frequencies is
motion of the ground, i.e. seismic noise.  Thermal motion of the
mirrors, their dielectric coatings and suspensions as well as
quantum radiation pressure noise are the other two types of
displacement noise which dominate in the low- to mid-frequency range
\cite{Saulson1990Thermal}. The primary sensing noise is shot noise
that arises from the Poisson statistics of photon arrival at the
photodetector.

\begin{figure}[Htb]
\centering
\includegraphics[width=0.8\textwidth]{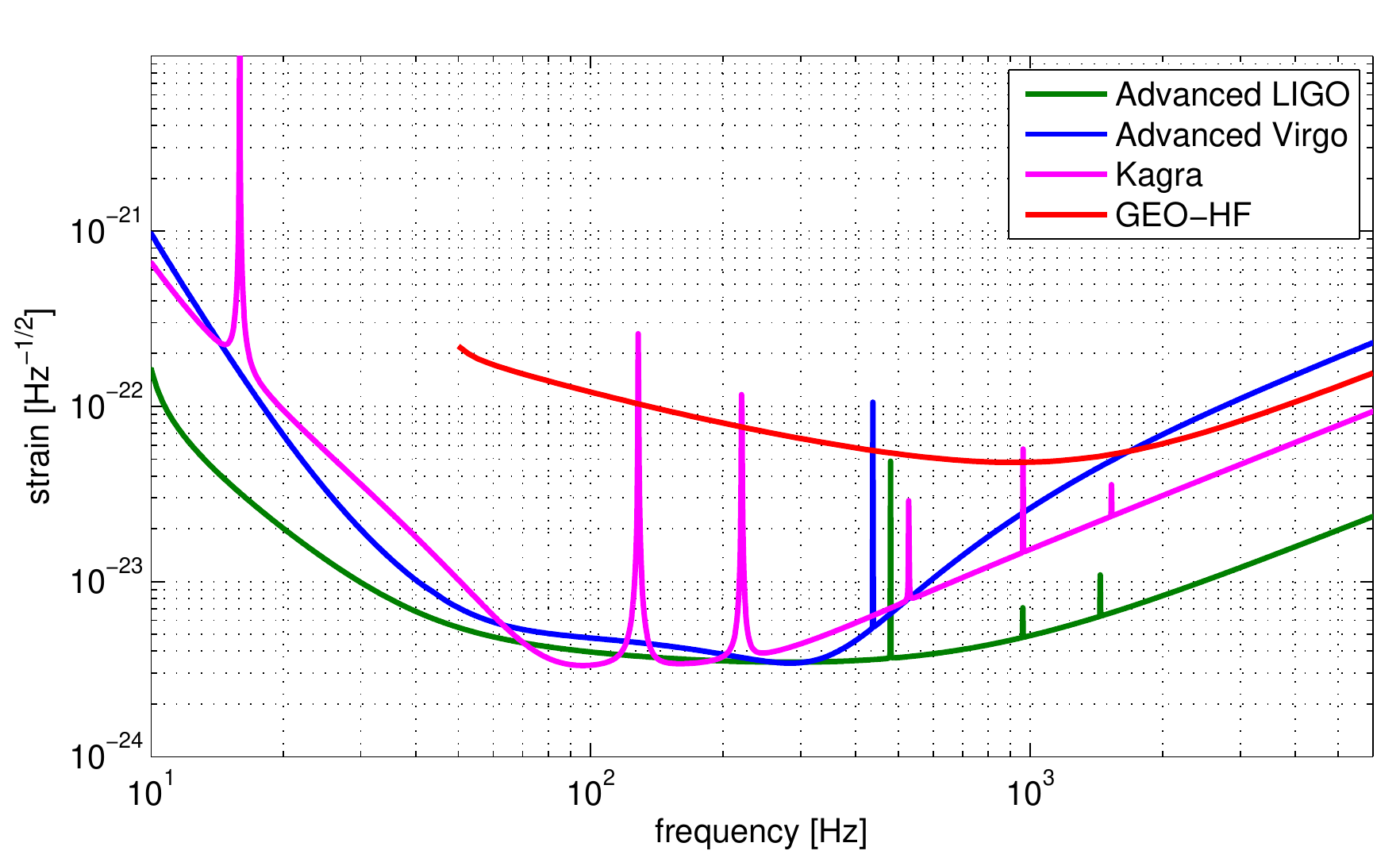}
\caption{Prediction of the incoherent sum of fundamental noise sources
  which give a likely best limit to the performance of each of the
  instruments in the advanced detector network. Noise curves for both
  Advanced Virgo and KAGRA show a de-tuned signal recycling
  configuration optimized for sensitivity to neutron star binaries
  \cite{AdVTDR, Somiya2012Detector, KAGRASensitivity}, whereas the
  Advanced LIGO and GEO-HF noise curves are shown with zero de-tuning
  \cite{LIGOSensitivity, GEOHFSensitivity}.}
\label{fig:designcurves}
\end{figure}

Figure \ref{fig:designcurves} shows the spectra of these ultimate
limits to the performance of each of the GW detectors in the advanced
detector network. Each curve reflects the incoherent sum of
fundamental noise sources, which gives a likely best limit to
performance. The actual sensitivity will depend also on technical
noise sources. The narrow lines in each of the noise curves represent
the thermally-excited violin modes of the test mass suspension
fibers. Quantum noise and thermal noise play the dominant roles in
limiting the sensitivity of each of the detectors. The frequency at
which there is a dip in the noise together with the shapes of the
noise curves are largely affected by the signal recycling parameters,
which may be adjusted during the course of operation of the advanced
detectors. We postpone the detailed description of each noise curve to
the proceedings dedicated to each detector. Table \ref{tab:overview}
provides an overview comparing some of the major properties of each of
the detector designs.

\begin{table*}
\centering
\caption{Some design properties of each of the four GWDs. DRFPMI
  stands for dual-recycled Fabry-Perot Michelson.}
\begin{tabular}{l l l l l}
\hline
           & Advanced LIGO & Advanced Virgo & GEO-HF & KAGRA \\
\hline
arm length & 4 km & 3 km & 2$\times$600 m & 3 km \\
power recycling gain & 44 & 39 & 900 & 11 \\
arm power  & 800~kW & 700~kW & 20~kW & 400~kW \\
\# of pendulum stages & 4 & 8 & 3 & 6 \\
mirror mass & 40~kg &42~kg & 6~kg & 23~kg \\
mirror material & fused silica & fused silica & fused silica & sapphire \\
temperature & room & room & room & cryogenic \\
topology & DRFPMI & DRFPMI & DRMI & DRFPMI \\
\hline
\end{tabular}
\label{tab:overview}
\end{table*}

\section{Outlook}
\label{summary}

The advanced detectors form a four-site network which is crucial for
GW signal characterization. The sky coverage depends on the detector
locations and orientations. The two LIGO detectors are nearly aligned
for maximum correlation, but they are relatively close together which
results in signals that have largely redundant information about the
source direction and character. For this reason, only the four-site
network can provide full sky coverage \cite{Schutz2011}.

Another important feature of a detector network is sky localization
for electro-magetic follow ups and multi-messenger investigations. In
the case of signals from coalescing binary neutron star systems, the
two Advanced LIGO detectors, Advanced Virgo, and KAGRA would provide a
localization accuracy of 10 $\mbox{deg}^2$ for $50\,\%$ of the
sources, while a five site network, including LIGO India, would
improve the accuracy to 5 $\mbox{deg}^2$ for $ 30\,\%$ of the
sources \cite{Fairhurst1,Fairhurst2}. This network of GW detectors
together with the joint detection of the electro-magnetic counterparts
is poised to open a promising window to gravitational wave astronomy.

\begin{figure}[Htb]
\centering
\includegraphics[width=0.8\textwidth]{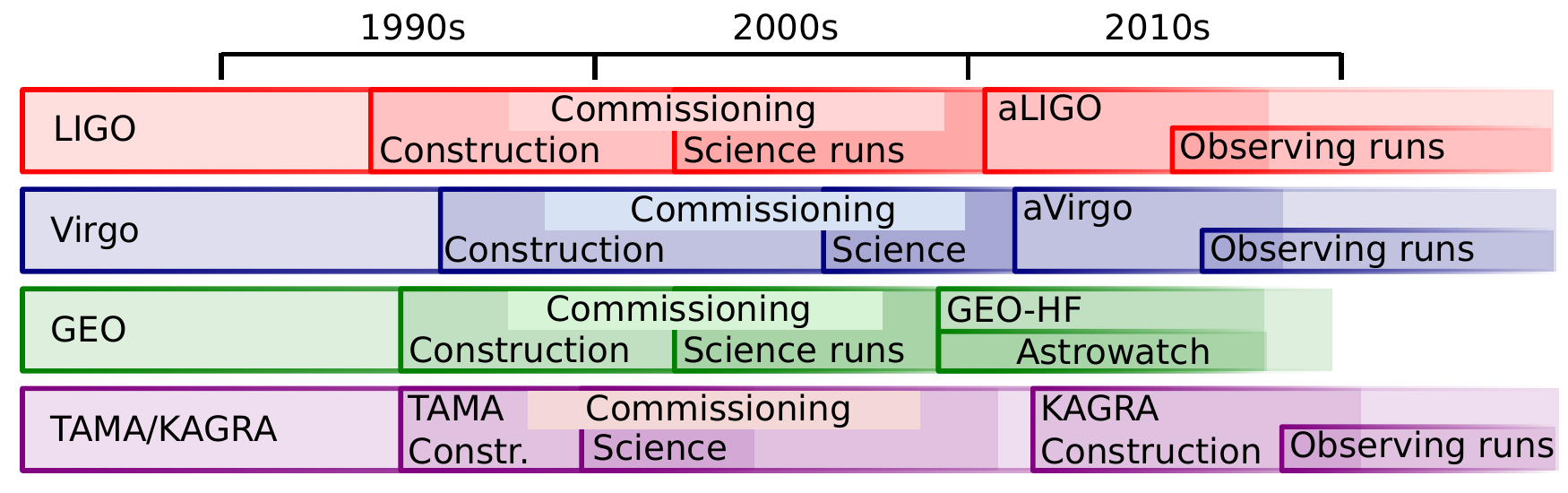}
\caption{Timeline showing the approximate dates for construction,
  commissioning and science runs of each of the ground based GWDs. The
  first joint science runs in the advanced detector era are expected
  to begin taking place by late 2015.}
\label{fig:timeline}
\end{figure}

Figure~\ref{fig:timeline} shows a timeline of the construction,
commissioning, and science run stages of each of the GWDs. The
down times for each of the detectors is a bit staggered, with GEO\,600
serving the role of being the sole detector online for the entirety of
the period when the other detectors are offline. Commissioning and
science run stages of the advanced detectors will be interspersed to
allow for possible early results once astrophysically interesting
sensitivities are reached, but before design sensitivity is
reached \cite{Collaboration2013Prospects}. The early science runs are
likely to start in late 2015 with the Advanced LIGO detectors, and
Advanced Virgo and KAGRA will join in due time.

Together with the promise of continuing to add to the collection of
upper limits on GW emission in the era of advanced detectors, the
first direct detection of GWs is expected to be made in only a couple
years' time from now. A first detection is expected to witness an
event such as a binary neutron star
coalescence \cite{Abadie2010Predictions}.

\section*{Acknowledgements}
The authors gratefully acknowledge the support of the United States
National Science Foundation for the construction and operation of the
LIGO Laboratory; the Science and Technology Facilities Council of the
United Kingdom, the Max-Planck-Society, and the State of
Niedersachsen/Germany for support of the construction and operation of
the GEO600 detector; the Italian Istituto Nazionale di Fisica Nucleare
and the French Centre National de la Recherche Scientifique for the
construction and operation of the Virgo detector; and the Japan
Society for the Promotion of Science (JSPS) Core-to-Core Program A,
Advanced Research Networks, and Grant-in-Aid for Specially Promoted
Research of the KAGRA project. This document has been assigned LIGO
document number LIGO-P1400153.

\section*{References}


\providecommand{\newblock}{}

\end{document}